\documentstyle[12pt]{article}
\begin{document}
\thispagestyle{empty}

\def\ve#1{\mid #1\rangle}
\def\vc#1{\langle #1\mid}

\newcommand{\p}[1]{(\ref{#1})}
\newcommand{\be}{\begin{equation}}
\newcommand{\ee}{\end{equation}}
\newcommand{\sect}[1]{\setcounter{equation}{0}\section{#1}}

\renewcommand{\theequation}{\thesection.\arabic{equation}}

\newcommand{\vs}[1]{\rule[- #1 mm]{0mm}{#1 mm}}
\newcommand{\hs}[1]{\hspace{#1mm}}
\newcommand{\mb}[1]{\hs{5}\mbox{#1}\hs{5}}
\newcommand{\Db}{{\overline D}}
\newcommand{\bea}{\begin{eqnarray}}
\newcommand{\eea}{\end{eqnarray}}
\newcommand{\wt}[1]{\widetilde{#1}}
\newcommand{\und}[1]{\underline{#1}}
\newcommand{\ov}[1]{\overline{#1}}
\newcommand{\sm}[2]{\frac{\mbox{\footnotesize #1}\vs{-2}}
           {\vs{-2}\mbox{\footnotesize #2}}}
\newcommand{\prt}{\partial}
\newcommand{\eps}{\epsilon}

\newcommand{\R}{\mbox{\rule{0.2mm}{2.8mm}\hspace{-1.5mm} R}}
\newcommand{\Z}{Z\hspace{-2mm}Z}

\newcommand{\cd}{{\cal D}}
\newcommand{\cg}{{\cal G}}
\newcommand{\ck}{{\cal K}}
\newcommand{\cw}{{\cal W}}

\newcommand{\vj}{\vec{J}}
\newcommand{\vl}{\vec{\lambda}}
\newcommand{\vz}{\vec{\sigma}}
\newcommand{\vt}{\vec{\tau}}
\newcommand{\vw}{\vec{W}}
\newcommand{\poiss}{\stackrel{\otimes}{,}}

\def\l#1#2{\raisebox{.2ex}{$\displaystyle
  \mathop{#1}^{{\scriptstyle #2}\rightarrow}$}}
\def\r#1#2{\raisebox{.2ex}{$\displaystyle
 \mathop{#1}^{\leftarrow {\scriptstyle #2}}$}}

\renewcommand{\thefootnote}{\fnsymbol{footnote}}
\newpage
\setcounter{page}{0}
\pagestyle{empty}

\vs{8}

\begin{center}

{\LARGE {\bf The invariant form of the generators of semisimple Lie and
quantum algebras in their arbitrary finite-dimensional
representation}}\\

\vs{8}

{\large  A.N. Leznov$^{a,2}$}
{}~\\
\quad \\
{\em {~$~^{(b)}$ Institute for High Energy Physics,}}\\
{\em 142284 Protvino, Moscow Region, Russia}\\
{\em {~$~^{(c)}$ Bogoliubov Laboratory of Theoretical Physics, JINR,}}\\
{\em 141980 Dubna, Moscow Region, Russia}

\end{center}

\vs{8}

\begin{abstract}

An explicit form  of the generators  of quantum and ordinary semisimple
algebras for
an arbitrary  finite-dimensional representation is found. The generators
corresponding to the simple roots are obtained in terms of a solution of a
system of
matrix equations. The result is presented in the form of  $N_l\times N_l$
matrices, where
$N_l$ is the dimension of the corresponding representation, determined by
the invariant
Weyl formula.

\end{abstract}

\vfill

{\em E-Mail:\
leznov@ce.ifisicam.unam.mx}

\newpage
\pagestyle{plain}

\renewcommand{\thefootnote}{\arabic{footnote}}

\setcounter{footnote}{0}

\section{Introduction}

In the present paper we pursue the main idea of the recent paper of the
author
\cite{1} (apart from some inessential technical details) to use the
character Weyl formula \cite{Weyl} for the construction in explicit form of
the
generators of both the quantum (and usual) semisimple algebras in an
arbitrary
finite-dimensional representation. In the construction of
\cite{1} the main idea was the proposition to use the Weyl character
formula for the
calculation of the result of the action of the group element $\exp \tau=
\exp \sum_1^r\tau_i h_i$ ($h_i$ are the Cartan elements of the algebra) on
the
basis state vectors of a finite-dimensional representation. This gives the
possibility
in the equations, defining a quantum algebra:
\begin{equation}
R_i X^{\pm}_j=\pm \tilde K_{j,i} X^{\pm}_jR_i,\quad
[X^+_i,X^-_j]=\delta_{j,i}
{R_i- R_i^{-1}\over 2 \sinh w_i t},\quad R_i\equiv \exp w_it h_i\label{1}
\end{equation}
to consider "group elements" $R_i$ in a given representation $l$ as  known
and
$X^{\pm}_i$ as the finite dimensional matrices with the known structure to
be
determined. In (\ref{1}) $K$ is the Cartan matrix,
$K_{j,i}w_i=w_jK_{j,i}\equiv
\tilde K_{j,i}$, $t$ is the deformation parameter.

{}From the first $2r^2$ equations of (\ref{1}) it is possible to obtain the
selection rules, which define the structure of the $X^{\pm}_i$ matrices.
The second $r^2$
equations define $X^{\pm}_i$ matrices uniquely up to an orthogonal
transformation, the sense of which will be explained below.

The present paper is devoted to a realization of the program described
above.

The paper is organized in the following way. In section 2 we describe the
connection between the Weyl character formula and and the result of the
action the
group element $\exp \tau$ on the basis vectors of an arbitrary
finite-dimensional
representation $l$. In section 3 the selection rules and explicit form of
the
equations which the "primitive" matrix elements of generators $X^{\pm}_i$
are satisfy are presented. The meaning  of the orthogonal transformation
and its
role in the construction is discussed in section 4. The consequences which
follow as the
result of  combining the selection rules with  orthogonal invariance are
discussed in the section 5. The explicit solution for generators
of each simple root in its canonical form is presented in the section 6.
In section 7 the main result in the form of the factorization theorem is
proved which restricts the solution of the whole problem to the case of the
simple algebras $A_2,B_2=C_2,G_2$. In section 8 the corresponding problem
is
solved for the algebras of the second rank. The general case of an
arbitrary
semisimple algebra is considered in section 9. Concluding remarks and
prospects for further consideration are gathered in section 10.

\section{ Weyl character formula and action of the group element $\exp
\tau$
on basis vectors of an arbitrary finite-dimensional representation}
Throughout this paper we make no distinction between the bases of the
quantum
and the usual semisimple algebras keeping in mind  that for their
finite-dimensional
representations, their dimensions and characters  (in the usual case) are
described by the famous Weyl formulae \cite{Weyl}.

The basis state vectors of a finite-dimensional irreducible representation
of a semisimple algebra $l=(l_1,l_2,..l_r)$ are constructed by repeated
application of the lowering $X^-_i$ generators to the highest vector
$\ve{l}$
with the properties:
$$
X^+_s \ve{l}=0,\quad h_s \ve{l}=l_s \ve{l}
$$
where $X^{\pm}_s,h_s$ are the generators corresponding to the simple roots
and
Cartan elements respectively.

The obvious difficulty of such a construction consists in the fact that not
all  state vectors arising in this way are linearly independent and an
additional
procedure for excluding linearly independent components with further
orthogonalization of the basis is necessary. Usually this is not a simple
problem.

Nevertheless, the values which the group element $\exp \tau\equiv \exp \sum
h_i\tau_i
$  takes on the basis vectors may be obtained from the invariant Weyl
character
formula for an irreducible representation $l=\sum h_il_i$.

The Weyl character formula:
\begin{equation}
\pi^l(\exp \tau)={ \sum_W \delta_W \exp (\tau_W, l+{1\over 2} \rho)\over
\sum_W \delta_W \exp (\tau_W, {1\over 2} \rho)}\label{WL}
\end{equation}
presented in form of the sum of exponents ( the denominator is always a
divisor of the numerator!):
$$
\pi^l(\exp \tau)=\sum_{n^k}^{N_l} C_{n^k} \exp (\tau, n^k)
=\sum_{n^k}^{N_l}
C_{n^k} \exp \sum_i^r (\tau_i n^k_i)
$$
gives the answer to many questions about the structure of the basis of the
corresponding representation. In (\ref{WL}) $W$ is the element of the
discrete Weyl group,
$\delta_W$ its signature, $\tau_W$ result of the action group element $W$
on
$\tau$, $C_{n^k}$ is the multiplicity of the corresponding exponent and
finally, ${1\over 2}\rho$ is half the sum of the positive roots of the
algebra.

 We remind the reader of the  definition:
$$
\pi^l(\exp \tau)= {\rm Trace} (\exp \tau)=\sum_{\alpha}^{N_l} \vc{\alpha}
(\exp \tau)
\ve{\alpha},
$$
where $\vc{\alpha},\ve{\alpha}$ are the bra and ket basis state vectors of
the
representation $l$.

Comparing the last two expressions we see that Weyl formula gives the
answer to
the question about the action of group element $\exp \tau$ on basis vectors
with
a given number of  lowering operators of different kinds. Indeed:
\begin{equation}
\exp \tau (X^-_1)^{m_1}...(X^-_r)^{m_r} \ve{l}=\exp \sum \tau_i l_i \exp
-\sum_{p,s} m_p\tilde K_{p,s} \tau_s (X^-_1)^{m_1}...(X^-_r)^{m_r}\ve{l}
\label{L}
\end{equation}
(of course the order of the lowering operators is inessential in the last
expression).

Equating each exponent of the Weyl formula to (\ref{L}) we can without any
difficulty find the indices $m_i$  corresponding to it. The only thing,
that
Weyl formula cannot do is to distinguish between  basis vectors
with the same number of  lowering operators of the same kind  taken in
different order. Nevertheless,  it defines the number of such states (i.e.
the multiplicity of the corresponding exponent in the formula for the
character)
These comments play a key role in the proposed construction.

\section{Selection rules and the system of equations for "primitive"
matrices}
Let us introduce generators $\bar {h^i}$  dual to $h_i$ with the
properties:
$$
[\bar {h^i}, X^{\pm}_j]=\pm \delta_{i,j} X^{\pm}_j,\quad {\rm Trace}(h_j,
\bar {h^i})=
\delta_{i,j}
$$
Generators $\bar {h^i}$ may be expressed as a linear combination of Cartan
elements:
$$
\bar h = h K^{-1}
$$
This fact can be checked by a simple direct computation.

The simplest and shortest way to calculate the values, which the group
element
$\exp (\bar h, p)=\exp \sum \bar {h^i} p_i$ takes on the basis vectors of
the finite-dimensional representation consists in  the exchange in the
formulae of the previous section  $\tau \to K^{-1} p$.

Let us now consider the family of the mutual commutative group elements
$$
\bar R_i=\exp \bar {h^i} t,\quad R_i=\prod_1^r (\bar R_s)^{\tilde K_{s,i}}.
$$
The eigenvalues of each such element and their multiplicity  can be
calculated
with the help of the Weyl character formula and have the form:
$$
C_{n^k} \exp t(n^kK^{-1})_i .
$$
Thus each basis vector of the finite-dimensional representation $l$ may be
marked with the help of the $r$-th dimensional vector
$q^k=(q^k_1,q^k_2,...q^k_r),
q^k_i=e^{(n^k K^{-1})_i}$. The multiplicity of the corresponding vectors
$q^k$ we denote by $N_k (\equiv C_{n^k})$.

Now let us consider the first system of $2r^2$ equations (\ref{1}).
Rewritten
in the terms of $\bar {R^i}$ generators, it takes the form:
\begin{equation}
\bar {R^i} X^{\pm}_j=\exp \pm \delta_{i,j} t X^{\pm}_j \bar
{R^i}\label{SR1}
\end{equation}
and means that the matrix elements of the matrices $(X^{\pm}_i)_{k,k'}$
are different from zero, when the indices $k,k'$ are connected by the
relation:
\begin{equation}
q^k_i=q^{k'}_i e^{\pm \delta_{i,j} t},\quad
\ln q^k_i-\ln q_i^{k'}=\pm(0,...t,...,0)\equiv \pm j \label{SR2}
\end{equation}
where the single element $t$  different from zero of the $j$th vector on
the left-
hand side sits in the $j$-th place. The $r$ -th dimensional vector
with  components $\ln q^k_i$ we denote by the single symbol $k$.

The last $r^2$ equations (\ref{1}) in the notation introduced above can be
rewritten as:
\begin{equation}
(X^+_i)_{k,k-i}(X^-_j)_{k-i,k-i+j}-(X^-_j)_{k,k+j}(X^+_i)_{k+j-i}=
\delta_{i,j} {\sinh \ln R^k_i\over \sinh w_i t} I_{N_k}\label{PMS}
\end{equation}
where $(X^{\pm}_i)_{k,k³}$ are rectangular $N_k\times N_{k'}$ matrices
which
we will call  "primitive" ones; $I_{N_k}$ is $N_k\times N_k$ unit matrix.

We emphasize once more that values of all components of vectors $q^k$ and
their corresponding multiplicity are known from the Weyl character
formula (\ref{WL}) and in the system (\ref{PMS}) must be considered as a
given. The unknowns are matrix elements of the rectangular primitive
matrices of
the given dimension.

\section{Orthogonal symmetry}

 It follows from the Weyl character formula (\ref{WL}) that the
 multiplicity of
each exponent in its involved will remain the same after reduction of the
group element $\exp \tau$ to an arbitrary subgroup of the initial
semisimple
group. This means, that all group elements $R_i,\bar {R^i}$ will have the
same
block structure -- the multiplicity of each exponent in the corresponding
place of each  will be the same. (This doesn't exclude the possibility of
the additional
degeneracy, when the eigenvalues of some of elements $R_i$ will be the same
for different diagonal blocks.) Thus each element $R_i$ is invariant with
 respect to similarity transformations generated by each $G(N_k,R)$
 subgroups.
The matrices $X^{\pm}_i$ preserve their form invariance, transforming in
accordance with the law:
$$
X^{\pm}_i\to G(N_k,R) X^{\pm}_i G^{-1}(N_k,R)
$$
or on the level of primitive matrices:
$$
(X^{\pm}_i)_{k,k'}\to G(N_k,R) (X^{\pm}_i)_{k,k'} G^{-1}(N_{k'},R)
$$

{}From the defining equations of the quantum algebra  (\ref{1}),
their invariance with respect to an inner automorphism follows:
$$
(X^-_i)^T\to X^+_i,\quad (X^+_i)^T\to X^-_i,\quad R^T_i=R_i
$$
The same is true also with respect to the system, which arises after taking
into
account the selection rules  which the primitive matrices  satisfy.

In what follows we will find generators of the simple roots of quantum
algebras
satisfying the additional conditions:
$$
(X^-_i)^T= X^+_i
$$
In this case all equations involved  preserve their invariance only after
reduction of the direct product of linear groups on a direct product of
orthogonal ones for which $O^{-1}=O^T$.

\section{The selection rules in combination with the orthogonal symmetry}

Let us consider separately the quantum algebra of the $j$-th simple root
with
corresponding generators $X^{\pm}_j, R_j$. The selection rules of  section
3
for generators $X^{\pm}_j$:
$$
\Delta \bar {h^i}=\pm \delta_{j,i}
$$
have as their direct corollary the corresponding selection rules for the
Cartan
elements $h_i$:
\begin{equation}
\Delta h_i=\pm K_{j,i} \label{SUPsr}
\end{equation}
since the matrix elements $(X^{\pm}_j)_{k,k'}$ are different from zero if
the
indices $k,k'$ are connected by the relation:
$$
k-k'=\pm j
$$
or in the language of the basis state vectors this means that the state
$\ve{k'}$ is distinguished from $\ve{k}$ by the action of exactly one
$X^-_j$ generator.
Provisionally it is possible to write this fact in symbolic form:
$$
\ve{k'}=[X^-_j \ve{k}]
$$
The quadratic brackets in the last relation mean that the state vector
$\ve{k³}$
is constructed as some linear combinations of the lowering generators of
the
basis state vector $\ve{k}$ by adding the  operator $X^-_j$ .

{}From (\ref{SUPsr}) it follows immediately that the generator
$(X^{\pm}_j)_{k,k'}$ is a direct sum of the generators of quantum $A^{q}_1$
algebras with known dimension of each component of this direct sum.
The deformation parameter for the quantum algebra connected with the $j$-th
simple root is equal to $w_j t$. So if we are able to present the explicit
expression for matrix elements $(X^{\pm}_j)_{k,k'}$ in the canonical form
for the $A^{q}_1$ algebra
then we can be sure that it is distinguished from the real one only by an
additional orthogonal transformation. If we would be able to find all
"mixing"
angles ( the orthogonal matrices of the corresponding dimension) of such
an
orthogonal transformation using equations from (\ref{PMS}) which connects
the generators of the nearest two or three  simple roots (in the cases of
the $D_n,E_{6,7,8}$
series), then this would give the  final solution to the problem.
In next few sections we consider a number of concrete examples and present
the solution for the general case of an arbitrary representation.

\section{Solution of $A^{q}_1$ equations in "canonical" form}

Let us choose from the sequence of the basis vectors, enumerated by the
$r$-th
dimensional vector $k$ those $k(p^s_i)$ ones, which are the highest weight
vectors
with respect to the representation $p^s_i$ of the quantum algebra $A^{q}_1$
($i$- is the number of  simple roots, $p^s_i$ is the index of its
$(2p^s_i+1)$ dimensional representation, $s$ - is the ordering number of
it). This means, that,
$$
(X^+_i)_{k(p^s_i)+i,k(p^s_i)}=0
$$
( in other words the state $k(p^s_i)+i$ is absent among the exponents of
the Weyl character formula).

In the mean time let us set aside all additional indices $k,i,s$
inessential for the problem of this section and consider the equations of
quantum algebra,
denoting its states by the usual $l, -l\leq m \leq l$ notation. We have:
\begin{equation}
X^+_{m,m-1}X^-_{m-1,m}-X^-_{m,m+1}X^+_{m+1,m}={\sinh 2mt\over \sinh t}
I_{N_m}
\label{QAE1}
\end{equation}
By $N_m$ we denote the dimension of the  vector with a given $m$.
For the state of the highest weight, ($X^+_{l+1,l}=0$), the corresponding
equation takes the form:
\begin{equation}
X^+_{l,l-1}X^-_{l-1,l}={\sinh 2lt\over \sinh t} I_{N_l} \label{QAE2}
\end{equation}
The only difference with respect to the usual quantum algebra case consists
in
the matrix character of $X^+_{l,l-1}$; in the usual case this is a single
c-number
function, in the case under consideration this is an $N_l\times N_{l-1}$
rectangular matrix. $N_l\leq N_{l-1}$ because at least $N_l$ basis states
of $N_{l-1}$ belong to the $(2l+1)$ multiplet. With the help of an
orthogonal
transformation $O_{N_l}$ from the left and $O_{N_{l-1}}$ from the right
the $N_l\times N_{l-1}$ rectangular matrix may be presented in the form
of an $N_l\times N_l$ square matrix with  elements different from zero on
its
main antidiagonal. From (\ref{QAE2}) all these elements are equal
to $X^+_{l,l-1}=({\sinh 2lt\over \sinh t})^{{1\over 2}}$. Thus the
remaining
$N_{l-1}-N_l$ basis vectors belong to $(l-1)$ representations of the
quantum
algebra as the highest weight vectors of these representations. By the same
reasoning,
$X^+_{l-1,l-2}$ is an $N_{l-1}\times N_{l-2}$ rectangular matrix, which
can be expressed as an antidiagonal $N_{l-1}\times N_{l-1}$ square matrix.
The first of its $N_{l-1}-N_l$ matrix elements (counting from the left
lower
corner) coincide with the matrix elements of quantum algebra  $A^q_1$;
$X^+_{l-1,l-2}=({\sinh 2(l-1)t\over \sinh t})^{{1\over 2}}$ in its $(l-1)$
representation. The next $N_l$ ones coincide with the  values
$X^+_{l-1,l-2}=({\sinh 2(l-1)t \sinh 2t\over \sinh^2 t})^{{1\over 2}}$,
the same elements but in the $l$ representation. The remaining elements
(if any)
 are the highest  weight vectors of the $(l-2)$-th representations and so
 on.

After the mid point  there arises only one difference
$N_{m+1}\leq N_m$ and everything is repeated in the opposite direction.
The matrix $X^+$ is symmetrical with respect to reflection in its
main antidiagonal.

In the case when multiplicities of all representations states are the same
$N_l=N_{l-1}=...=N_{-l}$ the canonical form of the generators $X^{\pm}$ is
a
similarity transform of the unit matrix of the corresponding dimension
multiplying
matrix elements of the scalar quantum $A^q_1$ algebra. This case will be
required
in section 9 for proving  the factorization theorem.

The  matrix $X^+$ constructed by above rules we will call  the canonical
one
and denote it by $\hat X^+$ with the all necessary indices which we have
omitted in the beginning of this section.

\section{Factorization of the problem}

 It follows from the results of the previous sections that the solution of
 the problem would be found if we would be able to find the orthogonal
 mixing angles
connecting all neighbouring dots (roots) on the Dynkin diagram of the
corresponding semisimple algebra. This problem in its turn can be achieved
by the same calculations in the case of algebras only  of the second rank.
Indeed let
$i$ and $i+1$ be adjacent dots on the Dynkin diagram. Let us reduce the
Weyl character formula on the subgroup of the second rank generated by the
simple roots. The result is as follows: the representation $l$ of the
initial semisimple
algebra is decomposed into the direct sum of irreducible representations of
$G^2_{i,i+1}$ ($i,i+1$ are the simple roots generating this  second rank
algebra). If we know the orthogonal mixing angles for
all $(p,q)$ representations of such groups of the second rank we can
resolve corresponding equations in the form  \footnote{ We write the
element
of orthogonal group only from the left omitting corresponding element
$O^{-1}$
from the right.}:
$$
X^+_i=\hat X^+_i,\quad X^+_{i+1}=\prod_{p,q} O^{(p,q)}(i,i+1) \hat
X^+_{i+1}
$$
where the product is taken over orthogonal mixing angles of all
representations $(p,q)$
into which the initial representation $l$ is decomposed under the
reduction described above.

Let us begin from the first simple root on the Dynkin diagram. We have
consequently
$$
X^+_1=\hat X^+_1,\quad X^+_2=\prod_{p,q} O^{(p,q)}(1,2) \hat X^+_2
$$
$$
X^+_2=\prod_{p,q} O^{(p,q)}(1,2) \hat X^+_2,\quad X^+_3=\prod_{p,q}
O^{(p,q)}
(1,2) \prod_{p³,q³} O^{(p³,q³)}(2,3) \hat X^+_3
$$
The next steps and the explicit form of the generators $X^+_i$ are
completely obvious.

Thus we reduce the solution of the problem with respect to an arbitrary
semisimple algebra to the same problem  with the respect to the algebras of
only
second rank. This assertion can be called the factorization theorem.
We will return to this question in section 9.

\section{ Algebras of the second rank}

In this case the basis vectors are labeled by a pair of indices ( the
vector
$k$ is two dimensional). We denote its components for the
highest weight vector by $\alpha,\beta.$
$$
\pmatrix{ \alpha \cr
          \beta \cr}= k^{-1}\pmatrix{ p \cr
                                      q \cr}
$$
where $(p,q)$ is the standard notation for  an irreducible representation
of the
of the second rank algebras $A_2,B_2=C_2,G_2$.
The highest weight vector has  unit multiplicity. All other basis
vectors may be decomposed on subspaces with a given number $N$ of lowering
generators. Each basis vector of this subspace may be denoted by the
natural
index $i$ ($1\leq i\leq N$) and "fine structure" index $m$ taking all
natural
values between its multiplicity (determined by the Weyl character formula)
and
unity. Thus for such a vector  we use the notation
$$
(\alpha-i,\beta-N+i)(m)
$$

The maximal available values for the multiplicity may be obtained from the
following consideration. Let us for the meantime set aside the problem of
linear
independent components. Then at each next step after application of the two
generators $X^-_{1,2}$ to each basis vector of the previous step the number
of
 basis vectors arising will be twice that for the previous step and thus
 the
maximal values of the multiplicity will be exactly the binomial
coefficients
$C^N_{\alpha-i,\beta-N+i}$.

Let us consider the first few lowest basis vectors in the meantime
forgetting about
their fine structure indices. We have the following chain:
$$
(\alpha,\beta),[(\alpha-1,\beta),(\alpha,\beta-1)],[(\alpha-2,\beta),
(\alpha-1,
\beta-1),(\alpha,\beta-2)],
$$
$$
[(\alpha-3,\beta),(\alpha-2,\beta-1),(\alpha-1,\beta-
2),(\alpha,\beta-3)],...
$$
where  the generators of the subspaces with the
given $N$ ($0,1,2,3,...)$  are gathered in square bracket.
The highest vector state $(\alpha,\beta)\equiv \ve{p,q}$ is annihilated by
the
both lowering generators and so is simultaneously the highest vectors of
the
$p$, $q$ representations of the $A_1(1,2)$ algebras generated respectively
by
the simple roots generators $X^{\pm}_{1,2}$. The state $X^-_2\ve{p,q}$ with
the
respect to the $A_1(1)$ algebra is the highest vector of the representation
$p-k_{21}
\equiv p+k$ ($k=1,2,3)$ and with the respect to $A_1(2)$ belongs to its
$q$-th
representation. The same situation takes place with respect to all other
basis
vectors and we will denote this fact by the additional upper, lower
indices describing the multiplet structure of the corresponding basis
states.

Thus the full notation which we use for the basis vectors states is the
following:
$$
(\alpha-i,\beta-N+i)^{p+N-i}_{q+i}(m)
$$

In this notation the first few first basis vectors take the form
$$
(\alpha,\beta)^p_q,[(\alpha-1,\beta)^p_{q+1},(\alpha,\beta-1)^{p+k}_q],
[(\alpha-2,\beta)^p_{q+2}, (\alpha-1,\beta-1)^{p+k}_{q+1},(\alpha,\beta-2)^
{p+2k}_q],
$$
$$
[(\alpha-3,\beta^p_{q+3},(\alpha-2,\beta-1)^{p+k}_{q+2},(\alpha-1,\beta-2)^
{p+2k}_{q+1},(\alpha,\beta-3)^{p+3k}_q],...
$$

After this preliminary comment we pass to concrete calculations.  After the
application  of the equation  $[X^+_1,X^-_2]=0$ to the basis vectors of the
first
subspace ($N=1$)  taking into account the selection rules of the
section 4 is equivalent to a single equation:
$$
(X^+_1)_{(\alpha,\beta-1),(\alpha-1,\beta-1)}(X^-_2)_{(\alpha-1,\beta-1),
(\alpha-1,\beta)}=
(X^-_2)_{(\alpha,\beta-1),(\alpha,\beta)}(X^+_1)_{(\alpha,\beta),(\alpha-1,
\beta)}
$$
Except for the  vector $(\alpha-1,\beta-1)$ the maximal multiplicity of
which is equal to $2$, all other  vectors involved are singlets and
so taking into account the definition of of the canonical form of the
quantum
algebra generators of the section 6 and the definition of orthogonal
matrices of
section 5 we rewrite the last equation:
$$
\sqrt{{\sinh (p+k)t\over \sinh t}}\pmatrix{1 & 0 \cr}
\pmatrix{ \cos \phi_1 & \sin \phi_1 \cr
         -\sin \phi_1 & \cos \phi_1 \cr}\sqrt{{\sinh (q+1)kt\over \sinh
         kt}}
         \pmatrix{ 1 \cr
                  0 \cr}=
\sqrt{{\sinh pt \sinh qkt\over \sinh t \sinh kt}}
$$
or
$$
\cos \phi_1=\sqrt{{\sinh pt \sinh qkt\over \sinh (p+k)t \sinh
(q+1)kt}},\quad
\sin \phi_1=\pm \sqrt{{\sinh ((p+qk+k)t \sinh kt \over \sinh (p+k)t
\sinh (q+1)kt}}
$$
Thus we have calculated the two (dimensional)   orthogonal  mixing  angle
and
multiplicity $2$ for the state vector $(\alpha-1,\beta-1)$. Only in the
case
(we assume $q\leq p$ but this absolutely inessential) $q=0$ this
multiplicity is degenerated up to  unity as  follows from the explicit 
expression for the mixing angle above. We emphasize that in the last 
calculations we have not used
Weyl formula but have calculated the multiplicity of the basis state
$(\alpha-1,\beta-1)$ independently.

For further calculations it will be more suitable to change notation for
the
basis vectors states and express them only in terms of the  basis vectors
of
two $A_{(1,2)}$ algebras. It is not difficult check that the basis vectors
of
the $N$-th subspace may be enumerator with the help only one index $0\leq
s\leq N$ as follows:
$$
((p+ks)_{N-s},(q+N-s)_s)
$$
In this notation the few first basis vector states have the form:
$$
(p_0,q_0),[(p_1,(q+1)_0),((p+k)_0,q_1)],[(p_2,(q+2)_0),((p+k)_1,(q+1)_1),
((p+2k)_0,q_2)],
$$
$$
[(p_3,(q+3)_0),((p+k)_2,(q+2)_1),((p+2k)_1,(q+1)_2),((p+3k)_0,q_3)]..
$$
$$
[(p_4,(q+4)_0),((p+k)_3,(q+3)_1),((p+2k)_2,(q+2)_2),((p+3k)_1,(q+1)_3),
((p+4k)_0,q_4)]..
$$
where $l_k$ is symbolically equivalent to $l_k\equiv (X^-)^k \ve{l}$ ( $l$
is
the index of representation of the $A_1$ algebra, $k$ its quantum number).

In this language the conditions of the the mutual commutativity of the
generators
$X^+_1$ and $X^-_2$ take the "factorizable" form ( generators
$X^{\pm}_{1,2}$
have matrix elements only between the states designated by indices $p$ and
$q$
respectively):
$$
(X^+_1)_{(p+ks)_{N-s},(p+ks)_{N-s+1}}(X^-_2)_{(q+N-s+1)_s,(q+N-s+1)_{s-1}}=
$$
\begin{equation}
(X^-_2)_{(q+N-s)_s,(q+N-s)_{s-1}}(X^+_1)_{(p+k(s-1))_{N-s},(p+k(s-1))_{N-s+
1}}
\label{ME}
\end{equation}

The last equations applied to three basis vectors of the $N=2$ subspace
lead to the pair of equations:
$$
(X^+_1)_{(p+k)_1,(p+k)_2}(X^-_2)_{(q+2)_1,(q+2)_0}=
(X^-_2)_{(q+1)_1,(q+1)_0}(X^+_1)_{p_1,p_2}
$$
$$
(X^+_1)_{(p+2k)_0,(p+2k)_1}(X^-_2)_{(q+1)_2,(q+1)_1}=
(X^-_2)_{q_2,q_1}(X^+_1)_{(p+k)_0,(p+k)_1}
$$
It is necessary to solve the last equations  under the additional
conditions that the multiplicity of the states $(p+k)_1,(q+1)_1$ is equal
to 2 and  the
maximal multiplicity of the states $(p+k)_2,(q+2)_1,(p+2k)_1,(q+1)_2$ are
is equal to 3. Using the notation $n((q+2)_1),\tilde n^2((q+1)_2)$ for
three-
dimensional orthogonal matrices we obtain for their matrix elements the
following values:
$$
n^1_1=-\sqrt {{\sinh (p-1)t \sinh (p+qk+k)t \sinh 2t \sinh kt\over
\sinh (p+k)t \sinh (p+k-2)t \sinh (q+2)kt \sinh t}}
$$
$$
n^1_2=\sqrt {{\sinh (p-1)t \sinh qkt \sinh pt \over
\sinh (p+k)t \sinh (p+k-1)t \sinh (q+2)kt}}
$$
$$
\tilde n^1_1=\sqrt {{\sinh ((p+qk+k)t \sinh 2kt\over
\sinh (p+2k)t \sinh (q+1)kt}},\quad
\tilde n^2_1=\sqrt {{\sinh pt \sinh (q-1)kt\over
\sinh (p+2k)t \sinh (q+1)kt}}
$$
{}From the last expression we see that in the first case we really have
"saturation" -- the multiplicity coincides with it maximal possible value,
except for the case of the $A_2$ algebra ($k=1$). In this case and in all
cases
 concerning the $\tilde n$ matrix we have:
$$
(n^1_1)^2+(n^1_2)^2=1,\quad (\tilde n^1_1)^2+(\tilde n^2_1)^2=1
$$
and we can assume that the multiplicity in these cases is equal to 2.
The equation for the real values of the multiplicities in these cases will
be
clarified after explicit calculations of all other matrix elements of the
$n,\tilde n$ matrices.
(We don't consider here the cases of the possible additional degeneracy
connected with particular values of the parameters $(p,q)$ defining the
representation).

In the case $N=3$ the main equations (\ref{ME}) take the form
$$
(X^+_1)_{(p+k)_2,(p+k)_3}(X^-_2)_{(q+3)_1,(q+3)_0}=
(X^-_2)_{(q+2)_1,(q+2)_0}(X^+_1)_{p_2,p_3}
$$
\begin{equation}
(X^+_1)_{(p+2k)_1,(p+2k)_2}(X^-_2)_{(q+2)_2,(q+2)_1}=
(X^-_2)_{(q+1)_2,(q+1)_1}(X^+_1)_{(p+k)_1,(p+k)_2}\label{ME3}
\end{equation}
$$
(X^+_1)_{(p+3k)_0,(p+3k)_1}(X^-_2)_{(q+1)_3,(q+1)_2}=
(X^-_2)_{q_3,q_2}(X^+_1)_{(p+2k)_0,(p+2k)_1}
$$

The multiplicities of the states $(p+2k)_1,(q+1)_2$ are equal to 2,
$(p+k)_2
(q+2)_1$ -- to 3, $(p+k)_3,(q+3)_1,(p+3k)_1,(q+1)_3$ -- not more than 4,
$(p+2k)_2,(q+2)_2$ -- not more than 6.

In terms of canonical forms of the generators $X^{\pm}_{1,2}$ from the
section
6 and orthogonal matrices of the sections 4,5 we rewrite the first equation
(\ref{ME3}):
$$
\pmatrix{ 0 & 0 & \sqrt {{\sinh (p+k-3)t \sinh 2t\over \sinh^2 t}} & 0 \cr
0 & \sqrt {{\sinh (p+k-2)t \sinh 3t\over \sinh^2 t}} & 0 & 0 \cr
\sqrt {{\sinh (p+k-4)t \over \sinh t}} & 0 & 0 & 0 \cr}
O_4 \sqrt {{\sinh (q+3)kt\over \sinh kt}}
\pmatrix{ 1 \cr
          0 \cr
          0 \cr
          0 \cr}=
$$
$$
n \sqrt {{\sinh (q+2)kt \over \sinh kt}}\pmatrix{ 1 \cr
                                                  0 \cr
                                                  0 \cr}\sqrt {{\sinh
                                                  (p-2)t
\sinh 3t \over \sinh^2 t}},
$$
where $O_4$ is a $4\times 4$ orthogonal matrix, $n$ a $3\times 3$
orthogonal matrix,
the components of which were partly calculated at the previous step.

The last equation allows to calculate only the matrix elements of the first
column of the $O_4$ matrix:
$$
O^1_1=\sqrt {{\sinh (q+2)kt \sinh (p-2)t \sinh 3t\over \sinh (q+3)kt
\sinh (p+k-4)t \sinh t}} n^1_3
$$
$$
O^1_2=\sqrt {{\sinh (q+2)kt \sinh (p-2)t \over \sinh (q+3)kt
\sinh (p+k-2)t}} n^1_2
$$
$$
O^1_3=\sqrt {{\sinh (q+2)kt \sinh (p-2)t \sinh 3t\over \sinh (q+3)kt
\sinh (p+k-3)t \sinh 2t}} n^1_1
$$
It is not difficult to check by direct computation that in the case of
$A_2$ algebra ($k=1$) the dimension of $O_4$ matrix is reduced  to $O_2$,
in the case of $B_2=C_2$ (k=2) it reduced  to $O_3$.

The second equation from (\ref{ME3}) with the same comments as in the
previous
case takes the form,
$$
\pmatrix{ 0 & \sqrt{{\sinh (p+2k-1)t \sinh 2t\over \sinh^2 t}} & 0 ...\cr
\sqrt{{\sinh (p+2k-2)t\over \sinh t}} & 0 & 0...\cr}P
\pmatrix{ 0 & 0 & \sqrt{{\sinh qkt\over \sinh kt}} \cr
          0 & \sqrt{{\sinh qkt\over \sinh kt}} & 0 \cr
\sqrt{{\sinh (q+1)kt \sinh 2kt\over \sinh^2 kt}} & 0 & 0 \cr}n^{-1}=
$$
$$
\tilde n
\pmatrix{ 0 & \sqrt{{\sinh (q-1)kt\over \sinh kt}} \cr
\sqrt{{\sinh qkt \sinh 2kt\over \sinh^2 kt}} & 0 \cr}
\phi^{-1}_1
\pmatrix{ 0 & \sqrt{{\sinh (p+k-1)t \sinh 2t\over \sinh^2 t}} & 0 \cr
\sqrt{{\sinh (p+k-2)t\over \sinh t}} & 0 & 0 \cr}
$$

In the last equations it is known that the dimension of the orthogonal
matrix $P$ is less than 6. But it is possible to determine only its 6
elements ( express
them in terms of matrix elements of $3\times 3$ orthogonal matrix $n$.
We present result of neither cumbersome  nor, on the other hand, very short
calculations in the form:
$$
\pmatrix{ 
\sqrt{{\sinh (q+1)kt \sinh 2kt\over \sinh^2 kt}} P^3_2 & \sqrt{{\sinh
qkt\over 
\sinh kt}} P^2_2 & \sqrt{{\sinh qkt \sinh 2t\over \sinh kt}} P^1_2 \cr 
\sqrt{{\sinh (q+1)kt \sinh 2kt\over \sinh^2 kt}} P^3_1 & \sqrt{{\sinh
qkt\over 
\sinh kt}} P^2_1 & \sqrt{{\sinh qkt \sinh 2t\over \sinh kt}} P^1_1 \cr}=
$$
$$
\pmatrix{ \sqrt{{\sinh (q-1)kt \sinh 2kt \sinh (p+k)t \sinh (p+k-1)t\over 
\sinh^2 kt \sinh (p+2k)t \sinh (p+2k-1)t}} & 0 \cr 
\sqrt{{\sinh pt \sinh (p+kq+k)t \sinh (p+k-1)t \sinh 2t\over \sinh t \sinh
(p+k)t
\sinh (p+2k)t \sinh (p+2k-2)t}} & -\sqrt{{\sinh qkt \sinh (p+2k)t
(p+k-2)t\over 
\sinh kt \sinh (p+k)t \sinh (p+2k-2)t}} \cr}\pmatrix{ n^1_2 & n^2_2 & n^3_2
\cr
                                                      n^1_1 & n^2_1 & n^3_1
                                                      \cr}
$$
In particular,
$$
P^3_2=\sqrt{{\sinh (q-1)kt \sinh (p+k)t \sinh (p+k-1)t\over 
\sinh (q+1)kt \sinh (p+2k)t \sinh (p+2k-1)t}}
$$
$$
P^3_1={\sinh 2kt\over \sinh t \sinh kt} \sqrt{{\sinh qkt \sinh 2t 
\sinh (p+qk+k)t \sinh (p-1)t\over \sinh (q+2)kt \sinh (p+2k)t}}
$$

The third equation from (\ref{ME3}) takes the form ( we have 
multiplied it addition
ally from the right on $2\times 2$ orthogonal matrix
$\tilde n$):
$$
\sqrt{{\sinh (p+3k)t\over \sinh t}} \pmatrix{ 1 & 0 & 0 & 0 \cr} R_4
\pmatrix{ 0 & \sqrt{{\sinh (q-4)kt \sinh 3kt \over \sinh^2 kt}} \cr
          \sqrt{{\sinh (q-2)kt \sinh 2kt \over \sinh^2 kt}} & 0 \cr
                                                          0 & 0 \cr
                                                          0 & 0 \cr}=
$$
$$
\sqrt{{\sinh (q-1)kt \sinh 3kt \over \sinh^2 kt}}\sqrt{{\sinh (p+2k)t\over
\sinh t}}\pmatrix{ 1 & 0 \cr}\tilde n
$$
In the last equation we have assumed the saturation case of the matrix $R$.
But
explicit calculation leads a the different result:
$$
R^2_1=\sqrt{{\sinh (p+qk+k)t \sinh 3kt \over \sinh (p+3k)t \sinh
(q+1)kt}},\quad
R^1_1=\sqrt{{\sinh (q-2)kt \sinh pt \over \sinh (p+3k)t \sinh (q+1)kt}}
$$
from which we can conclude that the multiplicity is equal to 2.

{}From the results of the present section we conclude that asymptotically
( the values of $(p,q)$ are sufficiently large) the multiplicities of the
spaces with
the different N are the following:
$$
N=1\to (1,1),\quad N=2\to (1,2,1),\quad N=3\to (1,3,2,1),\quad N=4\to
(1,4,4,2,1)
$$
and this result is obtained without using the Weyl character formula.
When the parameters of the representation reach some particular values the
multiplicity
changes by a jump, which is it possible to observe from the structure of
the
orthogonal matrix connected with the corresponding basis  vector.

\subsection{The case of $(2,1)$ representation of $B_2$ algebra}

{}From this concrete example the reader will be able to understand more
clearly
the main steps of the calculations and convinced himself of the self
consistency of the whole construction.

The dimension of the representation under consideration is equal to 35.
35 exponents of the Weyl character formula (the explicit expression for the
character
of the $(p,q)$ representation of the $B_2$ algebra  can be found
in Appendix I) may be distributed into subspaces with upper index $N$
($0\leq N\leq 10$) with the corresponding multiplicities in the following
order:
$$
[e^{(2\tau_1+\tau_2)}],[e^{\tau_2},e^{(4\tau_1-\tau_2)}],[e^{(-2\tau_1+
3\tau_2)},2e^{2\tau_1}],[3e^{\tau_2},e^{(4\tau_1-2\tau_2)}],[2e^{(-2\tau_1+
2
\tau_2)},3e^{(2\tau_1-\tau_2)}],
$$
$$
[e^{(-4\tau_1+3\tau_2)},3,e^{(4\tau_1-3\tau_2)}],
$$
$$
[3e^{(-2\tau_1+\tau_2)},e^{(2\tau_1-2\tau_2)}],[e^{(-4\tau_1+2\tau_2)},
3e^{-\tau_2}],[2e^{-2\tau_1},e^{(2\tau_1-3\tau_2)}],[e^{(4\tau_1-\tau_2)},
e^{-2\tau_2)}],[e^{(-2\tau_1-\tau_2)}]
$$
The same basis vectors in terms of irreducible representations of the $A_1$
algebra, distributed into the same subspaces appear as follows:
$$
[p_0,q_0],\quad [(p_1,(q+1)_0),((p+k)_0,q_1)],\quad [(p_2,(q+2)_0),
\phi_1((p+k)_1,(q+1)_1],\quad
$$
$$
[l((p+k)_3,(q+2)_1),((p+2k)_1,(q+1)_2)],\quad
[\phi_2(p+k)_3,(q+3)_1),m((p+2k)_2,(q+2)_2)],
$$
$$
[((p+k)_4,(q+4)_1),n((p+2k)_3,(q+3)_2),((p+3k)_2,(q+2)_3)],
$$
$$
[o((p+2k)_4,(q+4)_2),\phi_3((p+3k)_3,(q+3)_3)],\quad [((p+2k)_5,(q+5)_2),r
((p+3k)_4,(q+4)_3)],
$$
$$
[\phi_4((p+3k)_5,(q+5)_3),((p+4k)_4,(q+4)_4)],
[((p+3k)_6,(q+6)_3),((p+4k)_5,(q+5)_4)],[((p+4k)_6,(q+6)_4)]
$$
The notations $\phi_s$, $l,m,n,o,r$ mean 2-th, 3-th 
dimensional orthogonal matrices respectively, which is necessary to find.   

Of course in all expressions above it is necessary to put $p=2,q=1,k=2$.
We preserve the previous notation to give to the reader the possibility of
a
simpler comparison with the results of the previous section. The following
comment is necessary. Not all multiplets above begin from
the zero index. So if a multiplet begins for instance from the first index
$(p+k)_1$, this means that this is the multiplet with $l=2$ but not with
$l=4$
and so on.

The condition of the mutual commutativity of $X^{\pm}_1$ generators with
$X^{\mp}_2$, after equating to zero all matrix elements in the
basis constructed above ( it follows from the selection rules that the
products of generators
$X^+_1X^-_2$ and $X^-_2X^+_1$ conserve $N$ and so it is necessary to check
the matrix elements of the commutator with the fixed values of $N$) leads
to the a system of equations to be solved. Below we present this system in
abstract
form, its encoding with the help of the formulae of Appendix II and
explicit solutions
are obtained at each step of the calculation. (Of course all results
connected with
$N=1,2,3$ may be obtained with the help the direct substitution
$k=2,p=2,q=1$
into the  general formulae obtained in the previous section, but for the
fullness of the picture we present them once more).
$$
N=1
$$
$$
(X^+_1)_{(p+k)_0,(p+k)_1}\phi_2(X^-_2)_{(q+1)_1,(q+1)_0}=
(X^-_2)_{q_1,q_0}(X^+_1)_{p_0,p_1}
$$
$$
\sqrt{{\sinh 4t\over \sinh t}}\pmatrix{1 & 0 \cr}
\pmatrix{ \cos \phi_1 & \sin \phi_1 \cr
         -\sin \phi_1 & \cos \phi_1 \cr}\sqrt{{\sinh 4t\over \sinh 2t}}
         \pmatrix{ 1 \cr
                  0 \cr}=
\sqrt{{\sinh 2t\over \sinh t}}
$$
or
$$
\cos \phi_1={\sinh 2t\over \sinh 4t},\quad
\sin \phi_1=\pm {\sqrt{\sinh 6t \sinh 2t}\over \sinh 4t}
$$
$$
N=2
$$
$$
(X^+_1)_{(p+k)_1,(p+k)_2}(X^-_2)_{(q+2)_1,(q+2)_0}=
(X^-_2)_{(q+1)_1,(q+1)_0}(X^+_1)_{p_1,p_2}
$$
$$
\sqrt{{\sinh 2t\over \sinh t}}
\pmatrix{ 0 & \sqrt {{\sinh 3t \over \sinh t}} & 0  \cr
          1 & 0 & 0 \cr} l \sqrt{{\sinh 6t\over \sinh 2t}}\pmatrix{ 1 \cr
                                                                    0 \cr
                                                                    0 \cr}=
\phi_1\sqrt{{\sinh 4t\over \sinh 2t}}\pmatrix{ 1 \cr
                                               0 \cr}
\sqrt{{\sinh 2t\over \sinh t}}
$$
$$
l^1_1=\mp \sqrt{{\sinh 2t\over \sinh 4t}},\quad l^1_2=\sqrt{{\sinh t\sinh^2
2t
\over \sinh 3t\sinh 4t\sinh 6t}},\quad l^1_3=\sqrt{{\sinh 8t\sinh t\over
\sinh 3t\sinh 6t}}
$$
$$
N=3
$$
$$
(X^+_1)_{(p+2k)_1,(p+2k)_2}(X^-_2)_{(q+2)_2,(q+2)_1}=
(X^-_2)_{(q+1)_2,(q+1)_1}(X^+_1)_{(p+k)_1,(p+k)_2}
$$
$$
\sqrt {{\sinh 4t \over \sinh t}}\pmatrix{ 1 & 0 & 0 \cr} m
                          \pmatrix{ 0 & 0 & 1 \cr
                                    0 & 1 & 0 \cr
            {\sinh 4t \over \sinh 2t} & 0 & 0 \cr} l^{-1}=
\sqrt {{\sinh 4t \over \sinh 2t}}\pmatrix{ 1 & 0  \cr} \phi^{-1}_1
\sqrt {{\sinh 2t \over \sinh t}}
\pmatrix{ 0 & \sqrt {{\sinh 3t \over \sinh t}} & 0  \cr
          1 & 0 & 0 \cr},
$$
$$
m^3_1={\sinh 2t \over \sqrt {\sinh 4t \sinh 6t}}
$$
$$
m^2_1=(\cos \phi_1\sqrt {{\sinh 3t \over \sinh t}} l^2_2-\sin \phi_1
l^2_1),\quad
m^1_1=(\cos \phi_1\sqrt {{\sinh 3t \over \sinh t}} l^3_2-\sin \phi_1 l^3_1)
$$
$$
N=4
$$
$$
(X^+_1)_{(p+2k)_2,(p+2k)_3}(X^-_2)_{(q+3)_2,(q+3)_1}=
(X^-_2)_{(q+2)_2,(q+2)_1}(X^+_1)_{(p+k)_2,(p+k)_3}
$$
$$
\sqrt{{\sinh 2t\over \sinh t}}
\pmatrix{ 0 & 0 & \sqrt {{\sinh 3t \over \sinh t}} \cr
          0 & 1 & 0 \cr
          1 & 0 & 0 \cr} n \sqrt{{\sinh 4t\over \sinh 2t}}
\pmatrix{ 1 & 0 \cr
          0 & 1 \cr
          0 & 0 \cr} \phi^{-1}_2=
$$
$$
                        m \pmatrix{ 0 & 0 & 1 \cr
                                    0 & 1 & 0 \cr
            {\sinh 4t \over \sinh 2t} & 0 & 0 \cr} l^{-1}
\sqrt{{\sinh 2t\over \sinh t}}
\pmatrix{ 0 & 1 \cr
\sqrt {{\sinh 3t \over \sinh t}} & 0  \cr
                               0 & 0 \cr}
$$
In what follows the following notation will be convenient:
$$
\pmatrix{ N^1 & N^2 \cr}\equiv n \pmatrix{ 1 & 0 \cr
                                           0 & 1 \cr
                                            0 & 0 \cr} \phi^{-1}_2
$$
where $N^{1,2}$ are two three dimensional column vectors.
$$
N=5
$$
$$
(X^+_1)_{(p+2k)_3,(p+2k)_4}(X^-_2)_{(q+4)_2,(q+4)_1}=
(X^-_2)_{(q+3)_2,(q+3)_1}(X^+_1)_{(p+k)_3,(p+k)_4},
$$
$$
(X^+_1)_{(p+3k)_2,(p+3k)_3}(X^-_2)_{(q+3)_3,(q+3)_2}=
(X^-_2)_{(q+2)_3,(q+2)_2}(X^+_1)_{(p+2k)_2,(p+2k)_3}
$$
$$
\pmatrix{ 1 \cr
          0 \cr
          0 \cr}=\sqrt{{\sinh 4t\over \sinh 2t}} n
\pmatrix{ 1 & 0 \cr
          0 & 1 \cr
          0 & 0 \cr} \phi^{-1}_2 \sqrt{{\sinh 4t\over \sinh t}}\pmatrix{ 1
          \cr
                                                                         0
                                                                         \cr}
$$
$$
\sqrt{{\sinh 4t\over \sinh t}}\pmatrix{1 & 0 \cr}\phi_3
\sqrt{{\sinh 4t\over \sinh 2t}}
\pmatrix{ 0 & 1 & 0 \cr
          1 & 0 & 0 \cr} n^{-1}=
\pmatrix{ 0 & 0 & 1 \cr} m^{-1} \sqrt{{\sinh 2t\over \sinh t}}
\pmatrix{ 0 & 0 & \sqrt{{\sinh 3t\over \sinh t}} \cr
          0 & 1 & 0 \cr
          1 & 0 & 0 \cr}
$$
The resolution of the last equations is the following:
$$
{\sinh 4t\over \sqrt{\sinh 2t \sinh 6t}}\pmatrix{ N^1_1 & N^1_2 & N^1_3
\cr}=
\pmatrix{ o^1_3 & o^1_2 & \sqrt {{\sinh 3t\over \sinh t}}o^1_1 \cr},
$$
$$
{\sinh 4t\over \sqrt{\sinh 2t \sinh 6t}}\pmatrix{ R^2_1 & R^2_2 & R^2_3
\cr}=
\pmatrix{ m^3_3 & m^3_2 & \sqrt{{\sinh 3t\over \sinh t}}m^3_1 \cr}
$$
where  two three dimensional mutually orthogonal vectors $R^1,R^2$ are
defined by
the relation
$$
\pmatrix{ R^1 & R^2 \cr}=n \pmatrix{ 0 & 1 \cr
                                   1 & 0 \cr
                                   0 & 0 \cr}\phi^{-1}_3
$$
{}From the definition of $N^{1,2},R^{1,2}$ vectors the following equation
connects them:
$$
\pmatrix{ R^2 & R^1 \cr}=\pmatrix{ N^1 & N^2 \cr}(\phi_2+\phi_3)^{-1}
$$
$$
N=6
$$
$$
(X^+_1)_{(p+3k)_3,(p+3k)_4}(X^-_2)_{(q+4)_3,(q+4)_2}=
(X^-_2)_{(q+3)_3,(q+3)_2}(X^+_1)_{(p+2k)_3,(p+2k)_4}
$$
$$
\sqrt{{\sinh 2t\over \sinh t}}
\pmatrix{ 0 & \sqrt {{\sinh 3t \over \sinh t}} & 0  \cr
          1 & 0 & 0 \cr} r \pmatrix{ 0 & 0 & 1 \cr
                                    0 & 1 & 0 \cr
            {\sinh 4t \over \sinh 2t} & 0 & 0 \cr} o^{-1}=
$$
$$
\phi_3
\sqrt{{\sinh 4t\over \sinh 2t}}
\pmatrix{ 0 & 1 & 0 \cr
          1 & 0 & 0 \cr} n^{-1}\sqrt{{\sinh 2t\over \sinh t}}
\pmatrix{ 0 & 0 & 1 \cr
          0 & 1 & 0 \cr
\sqrt {{\sinh 3t \over \sinh t}} & 0 & 0 \cr} o \sqrt{{\sinh 6t\over \sinh
2t}}
$$
$$
N=7
$$
$$
(X^+_1)_{(p+3k)_4,(p+3k)_5}(X^-_2)_{(q+5)_3,(q+5)_2}=
(X^-_2)_{(q+4)_3,(q+4)_2}(X^+_1)_{(p+2k)_4,(p+2k)_5}
$$
$$
\sqrt{{\sinh 2t\over \sinh t}}
\pmatrix{ 0 & 1 \cr
\sqrt {{\sinh 3t \over \sinh t}} & 0  \cr
                               0 & 0 \cr}\phi_4\sqrt{{\sinh 4t\over \sinh
                               2t}}
         \pmatrix{ 1 \cr
                  0 \cr}=r \pmatrix{ 0 & 0 & 1 \cr
                                    0 & 1 & 0 \cr
            {\sinh 4t \over \sinh 2t} & 0 & 0 \cr} o^{-1} \sqrt{{\sinh
            4t\over
            \sin t}}\pmatrix{ 1 \cr
                              0 \cr
                              0 \cr}
$$
$$
N=8
$$
$$
(X^+_1)_{(p+4k)_4,(p+4k)_5}(X^-_2)_{(q+5)_4,(q+5)_3}=
(X^-_2)_{(q+4)_4,(q+4)_3}(X^+_1)_{(p+3k)_4,(p+3k)_5}
$$
$$
\sqrt{{\sinh 2t\over \sinh t}}\sqrt{{\sinh 4t\over \sinh 2t}}\pmatrix{1 & 0
\cr}
\phi^{-1}_4=\sqrt{{\sinh 6t\over \sinh 2t}} \pmatrix{ 0 & 0 & 1 \cr} r^{-1}
\sqrt{{\sinh 2t\over \sinh t}}
\pmatrix{ 0 & 1 \cr
\sqrt {{\sinh 3t \over \sinh t}} & 0 \cr
                               0 & 0 \cr}
$$
$$
r^3_1=\mp \sqrt{{\sinh 2t\over \sinh 4t}},\quad r^3_2=\sqrt{{\sinh t\sinh^2
2t
\over \sinh 3t\sinh 4t\sinh 6t}},\quad r^3_3=\sqrt{{\sinh 8t\sinh t\over
\sinh 3t\sinh 6t}}
$$
$$
N=9
$$
$$
(X^+_1)_{(p+4k)_5,(p+4k)_6}(X^-_2)_{(q+6)_4,(q+6)_3}=
(X^-_2)_{(q+5)_4,(q+5)_3}(X^+_1)_{(p+3k)_5,(p+3k)_6}
$$
$$
\sqrt{{\sinh 2t\over \sinh t}}=\sqrt{{\sinh 4t\over \sinh 2t}}
\pmatrix{1 & 0 \cr}
\pmatrix{ \cos \phi_4 & \sin \phi_4 \cr
         -\sin \phi_4 & \cos \phi_4 \cr}\sqrt{{\sinh 4t\over \sinh t}}
         \pmatrix{ 1 \cr
                  0 \cr}
$$
or
$$
\cos \phi_4={\sinh 2t\over \sinh 4t},\quad
\sin \phi_4=\pm {\sqrt{\sinh 6t \sinh 2t}\over \sinh 4t}
$$

It is more convenient to consider the system above grouping its equations
into the
pairs $(1,9),(2,8),(3,7),(4,6), (5,5)$.

{}From the first pair we find the two dimensional mixing angles
$\phi_1=\phi_4$.
Equations $(2,8)$ allow us to determine the first and the last columns of
the
 three dimensional orthogonal matrices $l$ and $r$ respectively and
 reconstruct
their explicit form up to arbitrary two dimensional rotations \footnote{
Equation $N=8$ after its transposition for the unknown function $rW$
coincides
with the same for the $l$ matrix. The same comments are true with respect
to all pairs enumerated above.} :
$$
l=\phi^l_{12} \phi^l_{13} u_{2,3}\equiv \bar l u_{2,3},\quad
rW=\phi^l_{12} \phi^l_{13} u_{2,3}\equiv \bar r W v_{2,3},
$$
where $W$ is a three dimensional matrix with elements different from zero;
ones on
its main antidiagonal and $g_{ij}$ are two dimensional rotation matrices in
the $(i,j)$ plane, parametrised by the parameter $g$;
$$
\sin \phi_{1,2}^l=-\sqrt{{\sinh 2t\sinh t\over \sinh 4t\sinh 5t}}\quad
\sin \phi^l_{1,3}=-\sqrt{{\sinh 8t\sinh t\over \sinh 3t\sinh 6t}},
$$
$u_{2,3},v_{2,3}$ two dimensional rotations in the $(2,3)$ plane with
arbitrary
parameters $u$ and $v$.

Equations $(N=3,N=7)$ give the possibility to find the first rows of the
matrices
$m$, $o$ and reconstruct them up to arbitrary two dimensional rotations:
$$
m=w_{2,3} \phi^m_{12} \phi^m_{13} u_{1,2}\equiv w_{2,3} \bar m
u_{1,2},\quad
oW=z_{2,3} \phi^m_{12} \phi^m_{13} v_{1,2}\equiv \bar o W v_{1,2},
$$
where
$$
\sin \phi_{1,2}^m=-\sqrt{{\sinh 6t\sinh t\over \sinh 4t\sinh 5t}}\quad
\tan \phi^m_{1,3}=-{\sinh 2t\over \sinh 4t}\sqrt{{\sinh 2t\sinh 5t\over
\sinh t
\sinh 8t}}
$$
and the parameters $(u,v)$ are the same as in the parametrisation of $l,r$
matrices and$(w,z)$ are arbitrary new ones.

Equations $(N=4,N=6)$ determine the pair of mutual orthogonal vectors
$(N^1,N^2)$,$(R^1,R^2)$ ($(N^1)^2=(N^2)^2=(R^1)^2=(R^2)^2=1, (N^1N^2)=
(R^1R^2)=0$).
The orthonormality conditions are not  additional equations but a
direct corollary of the previous equations for the matrices $l,r,m,o$ .

$$
(N^1,N^2)=\sqrt{{\sinh 2t\over \sinh 4t}} w_{1,2}
\pmatrix{ 0 & 0 & 1 \cr
          0 & 1 & 0 \cr
\sqrt {{\sinh 3t \over \sinh t}} & 0 & 0 \cr} \bar m
\pmatrix{ 0 & 0 & 1 \cr
          0 & 1 & 0 \cr
{\sinh 4t \over \sinh 2t} & 0 & 0 \cr} \bar l^{-1}
\pmatrix{ 0 & 1 \cr
\sqrt {{\sinh 3t \over \sinh t}} & 0 \cr
          0 & 0 \cr}\equiv w_{1,2}(\bar N^1,\bar N^2)
$$
$$
(R^1,R^2)=\sqrt{{\sinh 2t\over \sinh 4t}} z_{1,2}
\pmatrix{ 0 & 0 & 1 \cr
          0 & 1 & 0 \cr
\sqrt {{\sinh 3t \over \sinh t}} & 0 & 0 \cr} \bar o
\pmatrix{ 0 & 0 & 1 \cr
          0 & 1 & 0 \cr
{\sinh 4t \over \sinh 2t} & 0 & 0 \cr} \bar r^{-1}
\pmatrix{ 0 & 1 \cr
\sqrt {{\sinh 3t \over \sinh t}} & 0 \cr
          0 & 0 \cr}\equiv z_{1,2}(\bar R^1,\bar R^2)
$$
We pay attention of the reader that the equations defined the $(N,R)$
vectors
contain only two ( arbitrary up to now) parameters $(w,z)$.

The equations $N=5$ connect $N^1$ ($R^2$) vectors with the first (third)
columns
of the matrices $(o,m)$ respectively. We restrict ourselves to
consideration of the
first equation. The consequences of the second one are absolutely the same.

The vector $N^1$ and the first column of the matrix $o$ are fully defined
(as it follows
from the explicit formulae for them above) up to orthogonal transformations
$w_{1,2},z_{2,3}$ respectively ( rotation $v_{2,3}$, arising after using
the
equality $v_{1,2}W=Wv_{2,3}$, gives no output into first column of $o$
matrix).
With the help of the direct computation it is not difficult to convince
that
indeed:
$$
{\sinh 4t\over \sqrt{\sinh 2t \sinh 6t}} N^1_3=\sqrt{{\sinh 4t \sinh
3t\over
\sinh t\sinh 6t}}(\bar m^1_1\bar l^3_2+\bar m^2_1\bar l^2_2+{\sinh 4t\over
\sinh 2t}\bar m^3_1\bar l^1_2)=\sqrt{{\sinh 3t\over \sinh t}}o^1_1
$$
(all explicit values inside the relation to be proved are known) and
$$
{\sinh^2 4t\over \sinh 2t \sinh 6t}=(o^1_3)^2+(o^1_2)^2+{\sinh 3t\over
\sinh t}
(o^1_1)^2
$$
Thus among the three equations connected two three dimensional vectors
$N^1$ and $o^1$ only one equation is essential. From this equation we
obtain:
$$
\cos (w+z)={\bar o^1_3\bar N^1_1+\bar o^1_2\bar N^1_2\over (\bar N^1_1)^2+
(\bar N^1_2)^2}
$$
The knowledge of the vectors  $N,R$ give
$$
\cos (\phi_2+\phi_3)={\sqrt{\sinh 6t \sinh 2t}\over \sinh 4t}
$$
So we have determined all parameters which permit the reconstruction
 in explicit
form  of the infinitesimal generators of the simple roots of the quantum
$B_2$ algebra
in its $(2,1)$ representation.

It is not surprising that in some of the cases we were able to obtain
only some particular combination of parameters. Namely in this combination
they arise
in infinitesimal operators, which in the usual for quantum mechanical
terminology are the observables in the problem under consideration.
The reader can convinced from the corresponding formulae of the text of the
truth of this assertion..

\section{The general case of arbitrary semisimple (quantum) algebra}

Let us consider the state vector of a finite dimensional representation
$(l)$
of an arbitrary quantum algebra with a fixed number of the lowering
generators:
$$
(X^-_1)^{m_1} (X^-_2)^{m_2}....(X^-_r)^{m_r} \ve{(l_1,...l_r)}
$$
The order of the operators involved in the last formula is inessential.
After the action $(m_s+1)$ times on such a state vector by the generator
$X^+_s$ we
surely will obtain zero. But zero can arise at some intermediate step and
answer as to when
this question can be found after comparison  with the Weyl character
formula
(\ref{WL}) with (\ref{L}) as it was explained in section 1. Thus we can say
that
with respect to  the representation of the algebra $A_1(s)$ the state
vector above
belongs to its $(l_s-\sum m_tk_{ts})$-th finite dimensional irreducible
representation with the quantum number $m_s$. We would like emphasize once
more
that it may  happen that the first $j$   basis vectors of this
representation is forbidden by Weyl character formula. In this case the
index of the
representation is reduced to $jw_s$, where $w_s$ is the norm of the
$s$ - th simple root of the algebra under consideration.

Thus instead of a notation for basis vectors of the $l$ representation with
the help
of $r$-dimensional vectors $k$ (see section 5), it is possible and more
convenient to use  a notation connected with the indices and quantum
numbers of the irreducible
representations of $r$ formally independent $A_1(s)$ algebras.

{}From now on  we use the notation for basis vectors in the subspaces with
fixed $N$ ( the number
of the lowering generators):
$$
[...(l_s-\sum_{t\neq s} m_t K_{ts})_{m_s})...], \quad \sum m_s=N
$$

In connection with the equations defining a quantum algebra (\ref{1}), the
commutators of each pair $(i\neq j)$ of generators of positive ($i$),
negative ($j$)
simple roots are equal to zero. This condition written in the form of
matrix
elements of the corresponding commutator takes the form:
$$
(X^+_i)_{(l_i-\sum_{t\neq i} m_t K_{ti}-K_{ji})_{m_i-1}),(l_i-\sum_{t\neq
i} m_t
K_{ti}-K_{ji})_{m_i}} (X^-_j)_{(l_j-\sum_{t\neq j} m_t K_{tj})_{m_j+1},
(l_j-\sum_{t\neq j} m_tK_{tj})_{m_j}}=
$$
\begin{equation}
(X^-_j)_{(l_j-\sum_{t\neq j} m_t K_{tj}+K_{ij})_{m_j+1},
(l_j-\sum_{t\neq j} m_tK_{tj}+K_{ij})_{m_j}}(X^+_i)_{(l_i-\sum_{t\neq i}
m_t
K_{ti})_{m_i-1}),(l_i-\sum_{t\neq i} m_tK_{ti})_{m_i}}=0 \label{FFE}
\end{equation}

In fact this is exactly the equation (\ref{ME}) ( rewritten in another
notation with
$k=-K_{2,1}$,$1=-K_{1,2}$) used in all concrete calculations of the section
8.

Let us consider  first the equation (\ref{FFE}) in the case
$K_{i,j}=K_{j,i}=
0$, the case when two different simple roots are connected by zero matrix
elements of the Cartan matrix. We use the abbreviations $I=l_i-\sum_{t\neq
i} m_t
K_{ti}, J=l_j-\sum_{t\neq j} m_t K_{tj}$ and rewrite (\ref{FFE}):
$$
(X^+_i)_{I_{m_i-1},I_{m_i}}(X^-_j)_{J_{m_j+1},J_{m_j}}=(X^-_j)_{J_{m_j+1},
J_{m_j}}(X^+_i)_{I_{m_i-1},I_{m_i}}
$$
In these last relations the multiplicities of states $(I_{m_i},J_{m_j})$
are
equal ( these are only  different notations for the same state vector);
for the same reason multiplicities of each of the following pairs are equal
$(I_{m_i-1},J_{m_j+1})$,$(I_{m_i},J_{m_j+1})$,$(J_{m_j},I_{m_i-1})$.
This means that the matrix elements of $X^+_i,X^-_j$ are canonically
equivalent
under unitary transformation on matrix elements of the usual
one-dimensional quantum
algebra. See in this context the comments at the end of section 6.

Thus we conclude that the commutator of each pair of generators of
positive-negative
simple roots connected with zero valued elements of Cartan matrix are zero
as a direct corollary of the selection rules of  section 4.

So it is necessary   to solve equations of commutativity only for the pairs
of
simple root generators connected by the elements of the Cartan
matrix different from zero . In other words this  means that at each step
of calculations it
is necessary to find the orthogonal mixing angles between each pair marked
by
dots on the Dynkin diagram.

The technique of these calculations is described in the section 8. The
final
result is presented in the recurrence formulae of the section 7.

The considerations of this section may be considered as  proof of the main
assertion:
{\bf The problem of the explicit form of the  generators of simple roots of
semisimple
Lie and quantum algebras may be reduced to the solution of the same problem
for the second rank algebras }.

\section{Outlook}

The line of approach suggested in the  author's recent paper \cite{1}
has been transformed  in the present one into a closed mathematical scheme,
which allows us to obtain the
explicit form of the generators of semisimple Lie and quantum algebras in
their
arbitrary finite-dimensional representations.

It turns out that for this purpose it is necessary to introduce  new
objects
( to the best of our knowledge unknown up to now) - orthogonal
matrices, dimensions of which coincide with the multiplicities of
corresponding
exponents in the  Weyl character formula or, and this is equivalent, a
number of linear independent basis vectors of representation, which can be
constructed with a fixed number of the lowering operators of the same kind.

The calculation scheme of  section 8 ( see also examples of $(2,1), (1,1)$
representations of $A_1$ and $B_2$ algebras respectively, considered in
\cite{1})
allow us to obtain at every step the explicit expressions for certain
"primitive" matrices ( constructed with the help of orthogonal ones), from
which generators
of the corresponding simple roots are constructed. Subsequent application
of this
procedure to all dots of the Dynkin diagram solves the problem of finding
in
explicit form all  generators of simple roots of the algebra under
consideration  in its
arbitrary finite-dimensional representation ( see section 7 and 9).

The most surprising and intriguing result is the universal character of the
orthogonal
matrices in the asymptotic region of sufficient large representation
parameters. They do not depend on the basis of the representation dictated
by
Weyl character formula and can be calculated by independent means as  was
done
in section 8 for subspaces of basis vectors with $N=1,2,3,4$.

The "limiting values" calculated in this way  of orthogonal matrices have a
universal analytical
dependence (from matrix elements of Cartan matrix) for all algebras of the
second rank, repeating by their degeneration properties the structure of
corresponding discrete Weyl groups.

Observing the degeneracy properties of orthogonal matrices

the dimension
changes discretely when passing from one representation to another one )
it is possible to arrive at  definite conclusions about the structure of
the
representation bases equivalent to the corollaries which follow from
the Weyl character formula.

If it would it be possible to find independent arguments ( based on
representation or
structure theory of semisimple algebras) explaining the fact of the
existence and
main properties of the orthogonal matrices,  then the problem of the
explicit form of
generators would be solved with the help of corresponding formulae of
sections
7 and 9. In this case it may be turn out that all representation theory of
semisimple algebras is only a direct consequence of the corresponding
theory of
orthogonal matrices.

 We would like to finish this outlook by two little comments of a different
kind.

Firstly, there is no doubt that the construction of the present paper is
applicable without essential alteration to the case of semisimple Lie and
quantum
super-algebras. Its solution is connected only with correct manipulations
and
computation in the Grassman space.

Secondly, some reminders about the history of the problem.

The way proposed in the present paper ( see also \cite{1}) is diametrically
opposite to the usual used for solution of this problem beginning from the
famous I.M.Gel'fand and M.L.Tsetlin papers \cite{GZ}. The main idea of the
previous investigations  consisted in finding a family of mutually
commutative operators ( constructed from the generators of the algebra) the
eigenvalues of which are able to define the basis of the representation.
As the reader have seen no mention of such a family of operators (including
the
Casimir ones) was used in the proposed construction. Only facts from the
global representation theory of semisimple algebras encoded in the Weyl 
character formula lead to introduction of orthogonal matrices with the 
corresponding algorithm for their calculations.
In this the way of the present paper differs from previous investigations.

\noindent{\bf Acknowledgements.}

Author friendly thanks D.B.Fairlie and A.V.Razumov  for permanent
discussions 
in the process of working on this paper and great help in preparation the 
manuscript for publication.

This work was done under partial support of Russian Foundation of
Fundamental Researches (RFFI) GRANT-N 98-01-00330.

\section{Appendix I}

The character of $(p,q)$ representation of $B_2$ gives by the formula:
$$
\pi^{p,q}(\tau_1,\tau_2)={\sinh l_1\tilde \tau_1 \sinh l_2\tilde \tau_2-
\sinh l_1\tilde \tau_2 \sinh l_2\tilde \tau_1\over
\sinh 2\tilde \tau_1 \sinh \tilde \tau_2-\sinh \tilde \tau_2 \sinh \tilde
\tau_1}
$$
where $l_1-l_2=p+1,l_2=l_2,l_1=p+q+2,l_1+l_2=p+q+3,\tilde \tau_1=\tau_1,
\tilde \tau_2=\tau_2-\tau_1$. In the case under consideration $p=2,q=1$ the
last expression pass to:
$$
\pi^{2,1}(\tau_1,\tau_2)=16\cosh \tilde \tau_1\cosh \tilde \tau_2
(\cosh?2 \tilde \tau_1+\cosh \tilde \tau_1\cosh \tilde \tau_2+\cosh?2
\tilde \tau_2)-12\cosh \tilde \tau_1\cosh \tilde \tau_2-1
$$
The multiplication by terms of the last expressions leads to the sum of
exponents of Weyl formula presented in the corresponding place of the main
text.

\section{Appendix II}

Below we present the explicit canonical form of the generators
$X^{\pm}_{1,2}$
in the case of $(2,1)$ representation of $B_2$ algebra. In all formulae
$p=2,q=1,k=2$.
$$
(X^+_1)_{p_0,p_1}=\sqrt{{\sinh 2t\over \sinh t}},\quad
(X^+_1)_{p_1,p_2}=\sqrt{{\sinh 2t\over \sinh t}}
$$
$$
(X^+_1)_{(p+k)_0,(p+k)_1}=\sqrt{{\sinh 4t\over \sinh t}}\pmatrix{1 & 0
\cr},
\quad (X^+_1)_{(p+k)_1,(p+k)_2}=\sqrt{{\sinh 2t\over \sinh t}}
\pmatrix{ 0 & \sqrt {{\sinh 3t \over \sinh t}} & 0  \cr
          1 & 0 & 0 \cr},
$$
$$
(X^+_1)_{(p+k)_2,(p+k)_3}=
\sqrt{{\sinh 2t\over \sinh t}}
\pmatrix{ 0 & 1 \cr
\sqrt {{\sinh 3t \over \sinh t}} & 0  \cr
                               0 & 0 \cr},\quad
(X^+_1)_{(p+k)_3,(p+k)_4}=\sqrt{{\sinh 4t\over \sinh t}}\pmatrix{ 1 \cr
                                                                  0 \cr}
$$
$$
(X^+_1)_{(p+2k)_1,(p+2k)_2}=\sqrt{{\sinh 4t\over \sinh t}}
\pmatrix{1 & 0 & 0 \cr},\quad
(X^+_1)_{(p+2k)_2,(p+2k)_3}=\sqrt{{\sinh 2t\over \sinh t}}
\pmatrix{ 0 & 0  & \sqrt {{\sinh 3t \over \sinh t}} \cr
          0 & 1 & 0 \cr
          1 & 0 & 0 \cr},
$$
$$
(X^+_1)_{(p+2k)_3,(p+2k)_4}=\sqrt{{\sinh 2t\over \sinh t}}
\pmatrix{ 0 & 0 & 1 \cr
          0 & 1 & 0 \cr
\sqrt {{\sinh 3t \over \sinh t}} & 0 & 0 \cr},\quad
(X^+_1)_{(p+2k)_4,(p+2k)_5}=\sqrt{{\sinh 4t\over \sinh t}}
\pmatrix{1 \cr
         0 \cr
         0 \cr}
$$
$$
(X^+_1)_{(p+3k)_2,(p+3k)_3}=(X^+_1)_{(p+k)_0,(p+k)_1},\quad
(X^+_1)_{(p+3k)_3,(p+3k)_4}=(X^+_1)_{(p+k)_1,(p+k)_2},
$$
$$
(X^+_1)_{(p+3k)_4,(p+3k)_5}=(X^+_1)_{(p+k)_2,(p+k)_3},\quad
(X^+_1)_{(p+3k)_5,(p+3k)_6}=0(X^+_1)_{(p+k)_3,(p+k)_4},
$$
$$
(X^+_1)_{(p+4k)_4,(p+4k)_5}=\sqrt{{\sinh 2t\over \sinh t}},\quad
(X^+_1)_{(p+4k)_5,(p+4k)_6}=\sqrt{{\sinh 2t\over \sinh t}}
$$
$$
(X^+_1)_{q_1,q_0}=1
$$
$$
(X^-_2)_{(q+1)_1,(q+1)_0}=\sqrt{{\sinh 4t\over \sinh 2t}}\pmatrix{ 1 \cr
                                                                  0
                                                                  \cr},\quad
(X^-_2)_{(q+1)_2,(q+1)_1}=\sqrt{{\sinh 4t\over \sinh 2t}}\pmatrix{1 & 0
\cr},
$$
$$
(X^-_2)_{(q+2)_1,(q+2)_0}=\sqrt{{\sinh 6t\over \sinh 2t}}\pmatrix{ 1 \cr
                                                                   0 \cr
                                                                   0
                                                                   \cr},\quad
(X^-_2)_{(q+2)_2,(q+2)_1}=\pmatrix{ 0 & 0 & 1 \cr
                                    0 & 1 & 0 \cr
            {\sinh 4t \over \sinh 2t} & 0 & 0 \cr}
$$
$$
(X^-_2)_{(q+2)_3,(q+2)_2}=\sqrt{{\sinh 6t\over \sinh 2t}}
\pmatrix{ 0 & 0 & 1 \cr}
$$
$$
(X^-_2)_{(q+3)_2,(q+3)_1}=\sqrt{{\sinh 4t\over \sinh 2t}}
\pmatrix{ 1 & 0 \cr
          0 & 1 \cr
          0 & 0 \cr},\quad
(X^-_2)_{(q+3)_3,(q+3)_2}=\sqrt{{\sinh 4t\over \sinh 2t}}
\pmatrix{ 0 & 1 & 0 \cr
          1 & 0 & 0 \cr}
$$
$$
(X^-_2)_{(q+4)_2,(q+4)_1}=(X^-_2)_{(q+2)_1,(q+2)_0},\quad
(X^-_2)_{(q+4)_3,(q+4)_2}=(X^-_2)_{(q+2)_2,(q+2)_1},
$$
$$
(X^-_2)_{(q+4)_4,(q+4)_3}=(X^-_2)_{(q+2)_3,(q+2)_2}
$$
$$
(X^-_2)_{(q+5)_3,(q+5)_2}=(X^-_2)_{(q+1)_1,(q+1)_0},\quad
(X^-_2)_{(q+5)_4,(q+5)_3}=(X^-_2)_{(q+1)_2,(q+1)_1},
$$
$$
(X^-_2)_{(q+6)_4,(q+6)_3}=1
$$


\begin{thebibliography}{99}

\bibitem{1}

A.N.Leznov, {\bf New approach to representation theory of semisimple
Lie and quantum algebras}{}, (1999)

\bibitem{Weyl}

H. Weyl, {\bf Classical groups. Their invariants and representations}.
Princiton, Princeton University Press, 1946.

\bibitem{GZ}

I.M.Gel'fand and M.L.Tsetlin, {\bf Dokl. Acad. Nauk SSSR 71, 825-829,
(1950)}

I.M.Gel'fand and M.L.Tsetlin, {\bf Dokl. Acad. Nauk SSSR 71, 1017-1020,
(1950)}

\end{thebibliography}
\end{document}